\documentclass[%
  reprint,
superscriptaddress,
amsmath,amssymb,
pra,
]{revtex4-1}

\usepackage{graphicx}
\usepackage{dcolumn}

\usepackage{xcolor}
\usepackage{amsfonts}
\usepackage{cleveref}
\usepackage{ulem} 
\usepackage{capt-of}

\newcommand{\eql}[2]{\begin{equation}\label{#1} #2 \end{equation}}

\newcommand{\ali}[1]{\begin{align} #1 \end{align}}
\newcommand{\<}{\langle}
\renewcommand{\>}{\rangle}
\newcommand{\eref}[1]{(\ref{#1})}

\begin{document}

\title{Efficient stochastic simulation of rate equations and photon statistics of nanolasers}

\author{Emil C. Andr\'{e}}
\affiliation{Department of Photonics Engineering, Technical University of Denmark, {\O}rsteds Plads 345A, DK-2800~Kgs. Lyngby, Denmark}
\author{Jesper M{\o}rk}
\affiliation{Department of Photonics Engineering, Technical University of Denmark, {\O}rsteds Plads 345A, DK-2800~Kgs. Lyngby, Denmark}
\affiliation{NanoPhoton - Center for Nanophotonics, Technical University of Denmark, {\O}rsteds Plads 345A, DK-2800 Kgs. Lyngby, Denmark}
\author{Martijn Wubs}
 \email{mwubs@fotonik.dtu.dk}
\affiliation{Department of Photonics Engineering, Technical University of Denmark, {\O}rsteds Plads 345A, DK-2800~Kgs. Lyngby, Denmark}
\affiliation{NanoPhoton - Center for Nanophotonics, Technical University of Denmark, {\O}rsteds Plads 345A, DK-2800 Kgs. Lyngby, Denmark}

\begin{abstract}
Based on a rate equation model for single-mode two-level lasers, two algorithms for stochastically simulating the dynamics and steady-state behaviour of micro- and nanolasers are described in detail. Both methods lead to steady-state photon numbers and statistics characteristic of lasers, but one of the algorithms is shown to be significantly more efficient. This algorithm, known as Gillespie's First Reaction Method (FRM), gives up to a thousandfold reduction in computation time compared to earlier algorithms, while also circumventing numerical issues regarding time-increment size and ordering of events. The FRM is used to examine intra-cavity photon distributions, and it is found that the numerical results follow the analytics exactly. Finally, the FRM is applied to a set of slightly altered rate equations, and it is shown that both the analytical and numerical results exhibit features that are typically associated with the presence of strong inter-emitter correlations in nanolasers. 
\end{abstract}

\maketitle

\section{Introduction}
Optical cavities on the micro- and nanometer scale can reduce the number of available modes for light emission and increase the coupling of spontaneously emitted light into the cavity mode(s) \cite{Purcell1946}. This can be useful for laser devices, as it allows for low power consumption and high modulation speeds \cite{Yokoyama1989,Altug2006,Suhr2010,Ni2012}. These features, combined with the small footprint of the devices, make them promising candidates for on-chip optical interconnects \cite{Miller2017}, as well as for many other uses, like chemical and biochemical sensing and detection \cite{Gather2011,Shambat2013,Fikouras2018,Ma2019}. Therefore, micro- and nanolasers and their properties are currently a prominent subject of research.

One major area of interest in regards to nanolasers is the photon noise and photon statistics: With a relatively small number of intra-cavity photons and a large fraction of the spontaneously emitted photons ending up in the cavity mode(s), the associated quantum noise becomes increasingly important \cite{Gies2007, Chow2014, Kreinberg2017,Moelbjerg2013,Marconi2018}. Recently, stochastic methods have been used to simulate nanoscale lasers with large values of the spontaneous emission $\beta$-factor \cite{Lebreton2013,Puccioni2015,Mork2018}, and it was found that the noise and statistics of nanolasers can be captured surprisingly well by including only shot noise; that is, the noise associated with the discreteness of the photons and emitters in the cavity. 

In this work we will show that another stochastic approach, often used in the fields of chemistry and biochemistry \cite{Gillespie1976,Gillespie1977,Gibson2000,Pahle2009}, can be used to describe the laser dynamics, and can decrease the computation times by several orders of magnitude compared to previous approaches \cite{Lebreton2013,Puccioni2015,Mork2018}. Additionally, certain numerical assumptions made in \cite{Mork2018} can be avoided with the new method, leading to a robust and versatile stochastic approach with numerous applications. Here we use it to consider the intra-cavity photon probability density functions as a lasing system transitions from thermal to coherent emission, and we will investigate the prospects of introducing effectively altered rates of emission in the rate equations as a way to qualitatively describe the effects of collective inter-emitter correlations. 

The paper is structured as follows: In Section \ref{Sec:RE} we will introduce the basic laser model that we consider in most of the paper, and give a brief overview of the laser rate equations. In Section \ref{Sec:StochSim} we will review and discuss the stochastic simulation algorithm used in \cite{Mork2018} and then introduce another, more efficient algorithm for stochastically simulating the laser rate equations. Using the new algorithm, in Section \ref{Sec:PhotonDist} we will examine the intra-cavity photon distributions of the laser and in Section \ref{Sec:Asymmetry} we will examine the effects of introducing an effective asymmetry between the spontaneous and stimulated emission rates. 
Finally, we will summarize and discuss our results in Section \ref{Sec:Summary}.

\section{Laser model and rate equations}\label{Sec:RE}
\begin{figure}[t]
    \centering
    \includegraphics[width=0.35\textwidth]{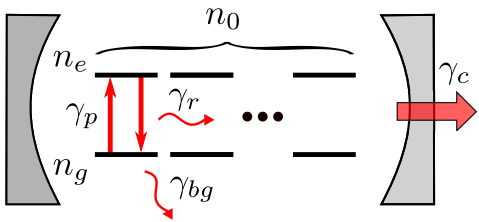}
    \captionof{figure}{Sketch of the two-level laser model.}
    \label{Fig:LaserModel}
\end{figure}

We consider a model for a nanolaser consisting of $n_0$ distinct two-level emitters interacting with a single cavity-mode, as shown schematically in Fig. \ref{Fig:LaserModel}. We assume for simplicity that all emitters couple to the cavity mode with the same coupling strength, so it is convenient to work with the total level populations $n_{e}$ and $n_{g}$ for the excited and ground levels, respectively. Since we consider two-level emitters, we have $n_{e} +n_{g}= n_0$. The intra-cavity photon number, i.e. the population of the cavity mode, will be denoted by $n_p$. The cavity mode has a decay rate per photon given by $\gamma_{c}$, and the emitters have a rate of loss into non-lasing modes given by $\gamma_{bg}$. The rate per emitter of emission into the lasing mode is denoted $\gamma_{r}$, giving a total decay rate $\gamma_{t} = \gamma_{r} + \gamma_{bg}$ from the excited state. The dependence of $\gamma_{r}$ on material and device parameters is discussed in \cite{Mork2018}, and includes Purcell enhancement. Finally, the emitters are pumped at a rate per emitter of $\gamma_{p}$. 

In this work we will focus on so-called Class B lasers, where the dephasing rate $\gamma_{2}$ of the emitters is much larger than the photon loss rate $\gamma_{c}$, so that the emitter polarization can be adiabatically eliminated from the analysis \cite{Arecchi1984}. For such devices, the laser dynamics may be described through rate equations for the photon number $n_{p}$ and the number of excited emitters $n_{e}$ \cite{Yokoyama1989, Bjork1991, Rice1994, Mork2018}:
    \ali{\dot{n}_{p} &= \gamma_{r} ( 2 n_{e} - n_{0} ) n_{p} + \gamma_{r} n_{e} - \gamma_{c} n_{p} + F_{p}, \label{RateEqs_p}\\
    \dot{n}_{e} &= \gamma_{p} ( n_{0} - n_{e} ) - \gamma_{r} ( 2 n_{e} - n_{0} ) n_{p} - \gamma_{t} n_{e} + F_{e}. \label{RateEqs_e}
    }
Here $F_{p}$ and $F_{e}$ are conventional zero-mean, delta-correlated stochastic Langevin terms, described explicitly in \cite{Mork2018,ColdrenCorzine}. They take into account the shot noise in the particle numbers due to the discrete nature of photons and excited emitters. The laser rate equations \eref{RateEqs_p}-\eref{RateEqs_e} give a simple and intuitively pleasing description of laser dynamics, since each term can be easily associated with a process occurring in the laser, as summarized in Table \ref{Table:Rates}. Note that the pumping term explicitly takes into account the effect of pump blocking \cite{Mork2018}.
The spontaneous emission $\beta$-factor is defined as the fraction of spontaneously emitted photons which end up in the cavity mode, i.e.
    \eql{beta-factor}{\beta = \frac{\gamma_{r} n_{e}}{\gamma_{t} n_{e}} = \frac{\gamma_{r}}{\gamma_{t}} = \frac{\gamma_{r}}{\gamma_{r} + \gamma_{bg}} . }

\begin{table}[t] 
    \centering
    \begin{tabular}{|l|l|l|}
    \hline
    \textbf{Event type} & $\pmb{a_{\mu}}$ & \textbf{Average rate}\\ \hline
    Stimulated Em. & $a_{st}$ & $\gamma_{r} n_{e}(t) n_{p}(t)$ \\ \hline
    Spontaneous Em. & $a_{sp}$ & $\gamma_{r} n_{e}(t)$  \\ \hline
    Absorption & $a_{a}$ & $\gamma_{r}(n_{0} - n_{e}(t)) n_{p}(t)$ \\ \hline
    Cavity decay & $a_{c}$ & $\gamma_{c} n_{p}(t)$\\ \hline
    Pumping & $a_{p}$ & $\gamma_{p} (n_{0} - n_{e}(t))$\\ \hline
    Background decay & $a_{bg}$ & $\gamma_{bg} n_{e}(t)$  \\ \hline
    \end{tabular}
    \caption{Average rates of event types in the laser rate equations.}
    \label{Table:Rates}
\end{table}

The rate equations \eref{RateEqs_p}-\eref{RateEqs_e} can be solved analytically in steady state, yielding expressions for the mean photon number $\<n_{p}\>$ and number $\<n_{e}\>$ of excited emitters as functions of the pump rate, $\gamma_{p} n_{0}$. These steady-state solutions can exhibit characteristic features of laser operation, namely emitter population clamping at large pump rates and a sudden jump in the photon number as a function of the pump rate. These features are characteristic of a system crossing the lasing threshold, but for certain parameter values, typically associated with $\beta\rightarrow 1$, these features become less distinct, and instead the photon statistics are used to characterise the transition to lasing \cite{Gies2007, Chow2014, Kreinberg2017,Moelbjerg2013}. The steady-state photon statistics of a laser are typically measured by the first two statistical moments of the photon number, or the mean value $\<n_{p}\>$ and the variance $\<\Delta n_{p}^2\> = \<n_{p}^2\> - \<n_{p}\>^2$. They are often summarised using the zero-delay second-order photon auto-correlation function $g^{(2)}(0)$ \cite{Loudon, ColdrenCorzine}. This can be expressed through the relative intensity noise (RIN) \cite{ColdrenCorzine} as 
    \eql{RIN_g2}{\text{RIN} = \frac{\<\Delta n_{p}^2\>}{\<n_{p}\>^2} , \hskip0.5cm g^{(2)}(0) = \frac{\<n_{p}(n_{p} - 1)\>}{\<n_{p}\>^2} = 1+ \text{RIN} - \frac{1}{\<n_{p}\>}.}
Lasing is then characterised by a transition from thermal (LED) light, for which $g^{(2)}(0) = 2$, to coherent (laser) light, for which $g{(2)}(0) \approx 1$.

\section{Stochastic Simulation}\label{Sec:StochSim}
In \cite{Mork2018} it was shown that stochastic simulations based on the algorithm presented in \cite{Puccioni2015} agreed very well with analytical results obtained from a small-signal analysis of the rate equations that is accurate even down to a few emitters ($n_{0} \geq 10$). In this section we will start by reviewing the specific method used for the stochastic laser dynamics simulations in \cite{Mork2018}, and then we will introduce a different method to produce the same results with significantly increased efficiency.

\subsection{Fixed Time-Increment Stochastic Algorithm}
The basic idea behind the stochastic simulations is to assume that the laser rate equations \eref{RateEqs_p}-\eref{RateEqs_e} describe a collection of discrete particles, namely photons and excited emitters, the numbers of which fluctuate due to several processes:  loss of particles (cavity or background losses), exchange of particles (emission or absorption) or addition of new particles (pumping). Individual events occur randomly, but their average rates are given by the terms on the right-hand sides of eqs. \eref{RateEqs_p}-\eref{RateEqs_e}, as summarized in Table \ref{Table:Rates}. 

In Appendix \ref{App:Algorithms}, we give a more detailed derivation of the simulation algorithm used in \cite{Mork2018}, but essentially it can be boiled down to the iteration over many time increments $dt$, where one asks the question ``how many events of type $\mu$ happened during $dt$?" for each event type: stimulated emission, absorption, spontaneous emission, cavity loss, background loss and pumping. In general, this is not an easy question to answer, since each event that occurs will immediately change the current number of photons and/or excited emitters; and this in turn affects the rates $a_{\mu}$ and the probability of subsequent events. However, in \cite{Mork2018} a few simplifying assumptions were made, which make it possible to approximate the answer: It was assumed that there is some order in which event types happen, in the sense that \emph{all} the events of a particular type occur together, before all events of the next type; it was assumed that one occurrence of an event does not affect the probability of any other events of the same type occurring; and it was assumed that the binomial distributions for the number of occurring events can be approximated by Poisson distributions. These assumptions are all reasonable when the value of $dt$ is sufficiently small. The algorithm is similar to the so-called $\tau$-leap method, described in detail in \cite{Pahle2009,Gillespie2001}.

Explicitly, the algorithm can be written as follows: \vskip2mm

\noindent
\textbf{Fixed Time-Increment Method for Laser Rate Equations (FTI)}
    \begin{enumerate}
        \item Initialize: Set system parameters, set initial number of photons $n_{p}(t=0)$ and excited emitters $n_{e}(t=0)$.
        \item Calculate rates $a_{\mu}$ for all event types $\mu$ according to Table \ref{Table:Rates} using current particle numbers.
        \item Determine the number of each type of event $\mu$ occuring during $dt$ by random draws from Poisson distributions with parameters $a_{\mu}dt$. 
        \item Update $n_{p}$ and $n_{e}$ accordingly, set $t \rightarrow t+dt$.
        \item If the maximal simulated time has been reached, end. Otherwise, go to step 2.
    \end{enumerate}

Solving the laser rate equations with this algorithm leads to equidistant time series for the number of intra-cavity photons and excited emitters, $(n_{p}(t_{i}) , n_{e}(t_{i}))$, and statistical convergence is ensured once reduction of the time increment $dt$ no longer affects the steady-state mean photon number $\<n_{p}\>$, excited-emitter number $\<n_{e}\>$, and the photon variance $\< \Delta n_{p}^{2} \>$. In practice, the size of $dt$ can be chosen at a given pump rate as some fraction $dt_{\rm{frac}}$ of the smallest reciprocal rate appearing in the rate equations \eref{RateEqs_p}-\eref{RateEqs_e}, computed using the analytical steady-state values for the particle numbers. For instance, if the mean steady-state rate of spontaneous emission is the largest among the steady-state rates in eqs. \eref{RateEqs_p}-\eref{RateEqs_e}, then $dt = dt_{\rm{frac}} / \gamma_{r}\<n_{e}\>$. This ensures that most iterations in the low-pump limit involve at most one event of any type, and convergence is typically obtained for $dt_{\rm{frac}}$ on the order of $10^{-2}$ or less. For all FTI simulations in this work we have used $dt_{\rm{frac}} = 10^{-2}$. 
The mean photon and emitter numbers obtained using the FTI show the behaviour expected for lasers, and the steady-state photon statistics, as computed using eqs. \eref{RIN_g2}, also exhibit the characteristic transition from chaotic to coherent light. This is shown in 
Fig. 3 of \cite{Mork2018}.

While the algorithm is stable and very easy to implement, it also has a few downsides. First, the computationally expensive act of drawing six Poisson-distributed random numbers at each iteration leads to long computation times, since many iterations are needed to obtain reliable photon statistics. Second, the assumption that all events of one type happen independently of all other event types introduces some ambiguity in the ordering of the events which happen during each time-increment: For instance, if there are $n_{e}(t)$ excited emitters at time $t$ and one is de-excited due to spontaneous emission into the lasing mode, then the number of excited emitters available for stimulated emission should in principle be $n_{e}(t)-1$. However, if the stimulated emission event occurred first, then the situation is opposite. This ordering issue may only play a small role in the outcome of the simulations as long as the time increment $dt$ is chosen sufficiently small, but this again leads to the question of how small is sufficient. All these challenges can be ameliorated by use of another stochastic simulation algorithm, which will be described in more detail below.

\subsection{Gillespie's First Reaction Method for the Laser Rate Equations}
\begin{figure*}[t]
    \centering
    \includegraphics[width=0.75\textwidth]{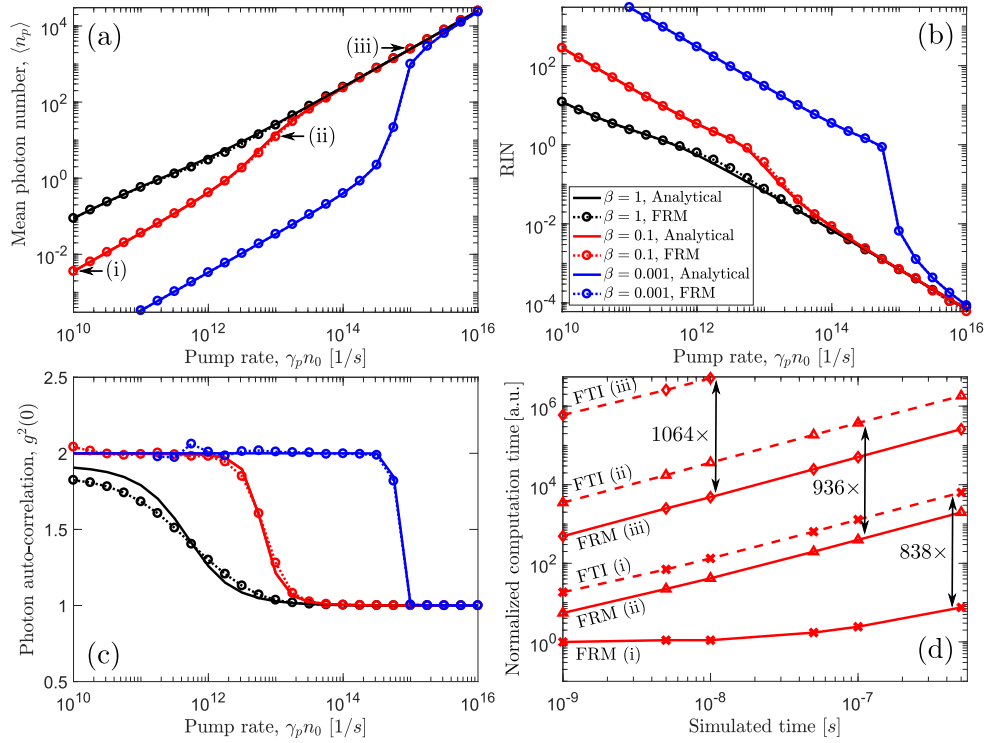}
    \caption{(a)-(c) Steady-state analytical [full lines] and numerical [markers] results for the mean photon number, relative intensity noise and photon auto-correlation versus pump rate. For all plots, the cavity decay rate is $\gamma_{c}~=~ 10^{11}\;s^{-1}$ and the total decay rate is  $\gamma_{t}=10^{10}\;s^{-1}$. The $\beta$-factor and number of emitters, $(\beta, n_{0})$, are varied between $(10^{-3},2\cdot 10^{4})$ [blue], $(10^{-1},2\cdot 10^{2})$ [red] and $(1,2\cdot 10^{1})$ [black]. For $\beta=10^{-3}$ [blue], low-pump results have been discarded because of statistical uncertainty. (d) Computation time using the FTI [dashed lines] and FRM [full lines] algorithms for different lengths of simulated time at the pump rates indicated in (a). Computation times have been normalized to the lowest computation time, which is 0.01~$s$, obtained using MatLab on a PC with a 2.7~GHz quad-core processor and 16~GB DDR4 RAM. The parameters are the same as those used for the $\beta=0.1$ plots in (a)-(c). 
    }
    \label{Fig:Simulation_SteadyState}
\end{figure*}

To obtain a faster algorithm, which has the added benefit of avoiding ordering issues as well as the need to define a time-step size, we can change the fundamental question asked at each iteration to "How long time until the next event occurs?" and "Which type of event will occur?". In this way, the size of the time increment at each iteration will differ, and only one event will happen during each time increment. Changing the basic point of view in this way corresponds to applying the algorithm known as Gillespie's First Reaction Method (FRM) \cite{Gillespie1976}, which is well-known in the chemistry and biochemistry communities \cite{Pahle2009}. It is used in numerical calculations regarding chemical reactions involving several species of molecules with finite populations, assuming that each type of reaction that can occur is a stochastic Markov process. The rate equations \eref{RateEqs_p}-\eref{RateEqs_e} describe an analogous type of system, where the photons and excited emitters are analogous to the particles, and the events of emission, absorption, loss and pumping are analogous to the reactions that change the particle populations. 

Changing the operational question of the simulation in this way means that instead of generating six integers corresponding to the change in particle numbers due to the six different event types, we are interested in generating two numbers: One corresponding to the time until the next event, and one corresponding to the type of event. As shown in \cite{Gillespie1976}, one way of doing this involves assuming for each of the different event types that it happens before any of the others, and then generating a tentative time $\tau_{\mu}$ until this event occurs. Once all the tentative times $\tau_{\mu}$ have been determined, we can choose the shortest, $\tau_{\mu_{0}}$, as the actual time until the first event (or reaction) occurs, and the corresponding type of event, $\mu_{0}$, as the event that occurs. The tentative time $\tau_{\mu}$ can be efficiently generated by drawing a random number from an exponential distribution with parameter $a_{\mu}$, which is proved in Appendix \ref{App:Algorithms} and in \cite{Gillespie1976}.

Specifically, the algorithm can be stated as follows:\vskip12mm

\noindent
\textbf{Gillespie's First Reaction Method for Laser Rate Equations (FRM)}
    \begin{enumerate}
        \item Initialize: Set system parameters, set initial number of photons $n_{p}(t=0)$ and excited emitters $n_{e}(t=0)$.%
        \item Calculate rates $a_{\mu}$ for all event types $\mu$ according to Table \ref{Table:Rates} using current particle numbers. %
        \item For each $\mu$, generate a tentative time increment $\tau_{\mu}$ by a random draw from an exponential distribution with parameter $a_{\mu}$. 
        \item Determine the event type $\mu_{0}$ for which $\tau_{\mu_0}~=~\min_{\mu}\{\tau_{\mu}\}$.
        \item Update $n_{p}$ and $n_{e}$ according to event type $\mu_0$, set $t \rightarrow t+\tau_{\mu_0}$.
        \item If the maximal simulated time has been reached, end. Otherwise go to step 2.
    \end{enumerate}

An important property of the FRM is that there are no numerical parameters, like the time increment $dt$ in the FTI algorithm. Therefore, there is no need for additional checks of convergence in terms of such parameters. In this sense, the FRM is an exact method to stochastically simulate the laser dynamics. In addition, exponentially distributed random numbers $\tau_{\mu}$ with parameters $a_{\mu}$ can be quickly and inexpensively generated using uniformly distributed random numbers $r_{\mu}$ on the unit interval $(0,1)$ as $\tau_{\mu} = -\log(r_{\mu})/a_{\mu}$ \cite{Robert2004,Krishnan}. Therefore, step 3 in the FRM involves six draws from a uniform distribution, instead of the six draws from six different Poisson distributions needed in the FTI algorithm. This drastically reduces the computation time of the FRM compared to the FTI, while the end results for the steady-state mean particle numbers and photon statistics are the same for both algorithms. This is shown in Fig. \ref{Fig:Simulation_SteadyState} by reproducing the results of Fig. 3 in \cite{Mork2018} through the use of the FRM, and comparing computation times versus simulated times for the two algorithms. We note that the very small mean photon numbers obtained for the blue curves in the low-pump range of Fig. \ref{Fig:Simulation_SteadyState} lead to relatively large statistical uncertainties. Therefore, results below a certain pump rate have been discarded, like in the corresponding figure of \cite{Mork2018}.

It is clear from Fig. \ref{Fig:Simulation_SteadyState}(d) that the computation time increases approximately linearly with the simulated time for both the FTI and FRM algorithms, as expected. However, as illustrated, the computation time at a fixed simulated time also increases for increasing pump rates. This is because the time delays between events become smaller at higher pump rates, as the larger particle numbers lead to larger rates of the different events of emission, absorption and loss. Smaller time increments imply more iterations needed to obtain the same simulated time, and hence larger computation times. In all cases, it is clear that the FRM algorithm leads to significantly shorter computation times at all pump rates: Indeed, improvements of several orders of magnitude can be obtained by using the FRM algorithm instead of the FTI algorithm. 

We see that the computation time for the FRM algorithm in the low-pump case is approximately constant for the shortest simulated times. This seeming lower bound on the computation time is related to the time for initialization rather than having to do with the specific algorithm. 

Since the FRM significantly reduces computation times, new features of the laser dynamics can easily be studied, which would otherwise be extremely time consuming. One such feature is the evolution of the photon number distributions within the laser cavity as the system transitions from chaotic, LED-like behaviour to coherent, laser-like behaviour. This will be examined in the next section.

\section{Photon distributions}\label{Sec:PhotonDist}

\begin{figure*}[t]
    \centering
    \includegraphics[width=0.75\textwidth]{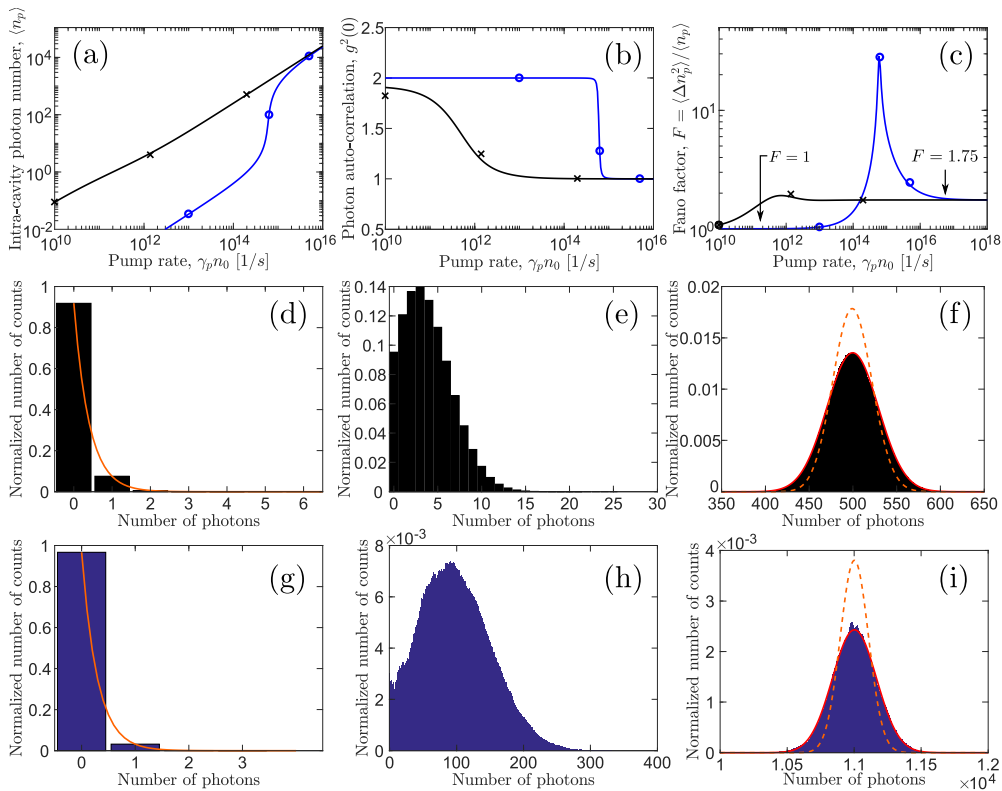}
    \caption{Mean photon number (a), photon auto-correlation (b) and Fano-factor (c) versus pump rate for a laser with $\beta=1$ [black curves] and $\beta=10^{-3}$ [blue curves]. (d)-(f) Normalised photon number distribution for the laser with $\beta=1$ at the pump values indicated with black crosses in (a)-(c). (g)-(i) Normalised photon number distribution for the laser with $\beta=10^{-3}$ at the pump values indicated with blue circles in (a)-(c). 
    The parameters used to make these plots are the same as those used to produce the curves in Fig. \ref{Fig:Simulation_SteadyState}. In (d,g) the orange curves indicate Bose-Einstein distributions with mean values determined by the mean photon numbers $\<n_{p}\>$. In (f,i) the orange dashed curves indicate Poisson distributions with parameters given by the mean photon number $\<n_{p}\>$, and the red curves indicate Gaussian distributions with means and variances given by the mean photon number $\<n_{p}\>$ and photon number variance $\<\Delta n_{p}^2\>$, respectively.
}
    \label{Fig:PhotonDist}
\end{figure*}
Once the laser settles into steady-state operation, the mean values for observables no longer change, but the particle numbers still fluctuate in time. By examining the simulated time series of the intra-cavity photon number in steady state, we can obtain the discrete photon probability density function (PDF) by evaluating how much of the total simulation time is spent with 1 photon in the cavity, how much is spent with 2 photons in the cavity, etc. In Fig. \ref{Fig:PhotonDist} the photon PDF obtained from the stochastic simulations is shown for three different values of the pump rate using parameters corresponding to the cases of $\beta=1$ and $\beta=10^{-3}$ from Fig. \ref{Fig:Simulation_SteadyState}. In both cases there is a clear transition from thermal statistics giving a Bose-Einstein distribution (d,g), through an intermediate distribution near threshold (e,h), to near-Poissonian statistics (f,i). 

As shown in Fig. \ref{Fig:PhotonDist}(b), the photon auto-correlation $g^{(2)}(0)$ seems to indicate that Poissonian statistics are obtained at large pump rates for both $\beta=1$ and $\beta = 10^{-3}$, the simulated values being $g^{(2)}(0) = 1.0015$ and $g^{(2)}(0)=1.0001$, respectively. However, we see in (f,i) that at high pump rates, the photon PDF approaches a Gaussian distribution which is wider than a Poisson distribution with the same mean value. Mathematically, we can explain this by the photon variance approaching a value slightly different from the characteristic Poissonian $\< \Delta n_{p}^{2} \> = \< n_{p} \>$. Equivalently, the Fano factor $ F = \<\Delta n_{p}^{2} \> / \< n_{p} \>$ does not reach a value of 1 at high pump rates. From the analytical expression in \cite{Mork2018} for the photon variance we can show that at large values of the mean photon number we have
    \eql{Fano_limit}{F = \frac{\<\Delta n_{p}^{2} \>}{\< n_{p} \>} \approx 1 + c, \hskip0.5cm \text{for } \<n_{p} \> \gg 1,}
where
    \eql{Fano_c}{c=\frac{1 + n_{th}/n_{0}}{2(n_{0}/n_{th} - 1)}.}
Here $n_{th} \equiv \gamma_{c} / \gamma_{r}$ is the semi-classical threshold value of the emitter population inversion. The result in eq. \eref{Fano_limit} is illustrated in Fig. \ref{Fig:PhotonDist}(c), where we have plotted the Fano factor versus the pump rate, with the constant $c=0.75$ for both values of $\beta$. Physically speaking, this deviation from exact Poissonian statistics has its roots in the laser level-scheme: Similar, non-Poissonian photon variances for lasers above threshold have been reported in earlier theoretical and experimental work \cite{Kozlovskii1994, Kozlovskii2014, vanDruten2000}, where two-, three- and four-level lasers are considered. In particular it is argued in \cite{Rice1994, Kozlovskii1994, Kozlovskii2014} that the two-level emitter scheme we consider here should generally lead to super-Poissonian statistics above threshold due to depletion of the emitter ground states. These effects are what we see in our analytical and numerical results. We note that the effects occur both when using the FTI and the FRM algorithms.

With the FRM algorithm it is possible to obtain photon distributions which match the analytical predictions almost perfectly, while the computation time is kept relatively short -- less than 2 hours on a commercially available laptop. This demonstrates once more the extreme efficiency and exactness of the FRM applied to the laser rate equations. In the next section we will use this to show how small alterations in the laser rate equations can lead to results which exhibit features that are typically obtained from much more intricate laser models. 

\section{Breaking the symmetry between spontaneous and stimulated emission rates}\label{Sec:Asymmetry}
In the traditional rate-equation description of the laser dynamics, the rate per emitter of spontaneous emission into the cavity mode is the same as the rate per emitter per photon of stimulated emission into the cavity mode, consistent with the Einstein relations \cite{ColdrenCorzine}. This can be seen in eqs. \eref{RateEqs_p}-\eref{RateEqs_e}, where the terms related to spontaneous and stimulated emission and absorption are all proportional to the rate of radiative decay, $\gamma_{r}$. However, it is possible to imagine that certain physical effects, which are not accounted for in the traditional derivation of the rate equations, could give rise to an effective difference between the rates of spontaneous and stimulated emission. For instance, collective effects due to correlations building up between emitters can affect the dynamics and steady-state behaviour of lasers, as has been shown in recent theoretical and experimental works \cite{Andre2019,Bohnet2012, Meiser2009, Meiser2010, Leymann2015, Jahnke2016, Kreinberg2017}. Some of these phenomena could potentially be captured by an asymmetry between the rates of spontaneous and stimulated emission, which is what we will investigate next.

To examine the effects of introducing a difference between the spontaneous and stimulated radiative events, we can replace the common rate $\gamma_{r}$  in eqs. \eref{RateEqs_p}-\eref{RateEqs_e} by two separate rates, $\gamma_{r}^{\rm sp}$ and $\gamma_{r}^{\rm st}$, for the spontaneous and stimulated events, respectively. The ratio 
    \eql{Eq_xi}{\xi = \gamma_{r}^{\rm sp} / \gamma_{r}^{\rm st}} 
quantifies the difference between the rates of spontaneous and stimulated emission (and absorption), and the rate equations \eref{RateEqs_p}-\eref{RateEqs_e} can be rewritten using this quantity. 
We can solve the rate equations including $\xi$ in the same way as we did for the traditional rate equations \eref{RateEqs_p}-\eref{RateEqs_e}, both analytically and numerically. If we perform the same small-signal analysis applied in \cite{Mork2018} with the new radiative rates, we can also obtain corresponding $\xi$-dependent analytical expressions for $\< \Delta n_{p}^{2} \>$, RIN and  $g^{(2)}(0)$. Incorporating the asymmetry in the stochastic simulations using the FRM is simple, since we only need to change the rates $a_{\mu}$ to include $\xi$. 
\begin{figure*}[t]
    \centering
    \includegraphics[width=0.68\textwidth]{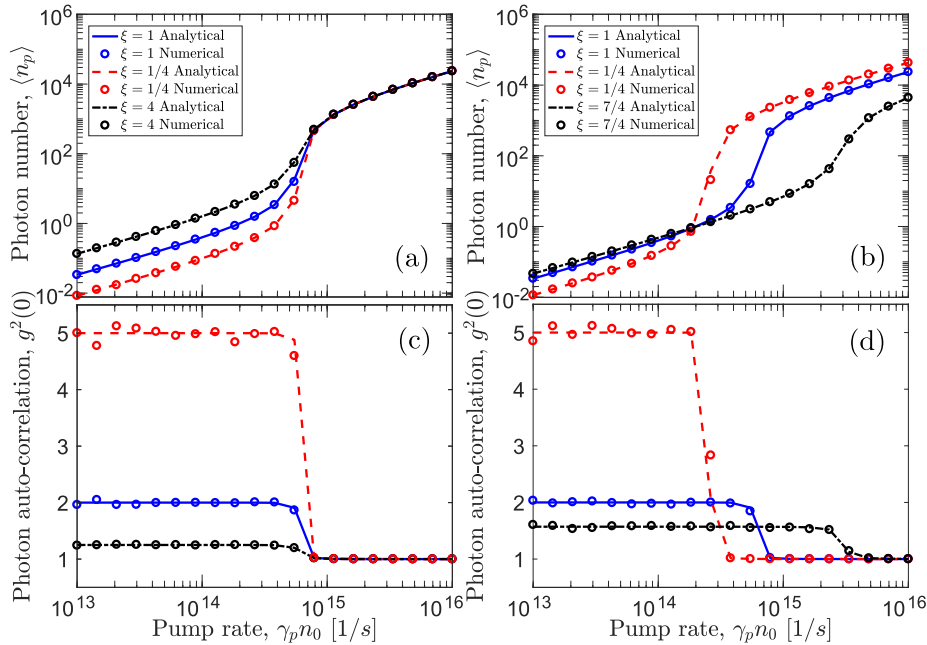}
    \caption{Photon number (a,b) and photon auto-correlation function (c,d) as functions of the pump rate for different values of the ratio $\xi= \gamma_{r}^{\rm sp}/\gamma_{r}^{\rm st}$. All plots are produced using $n_{0}=20000$. In (a,c) $\gamma_{r}^{\rm st}$ and hence $n_{th}= \gamma_{c} / \gamma_{r}^{\rm st}=10000$ are kept fixed, while $\gamma_{r}^{\rm sp}$ changes with $\xi$. In (b,d) $\gamma_{r}^{\rm sp}$ and hence $\beta = 10^{-3}$ are kept fixed, while $\gamma_{r}^{\rm st}$ and hence $n_{th}$ change with $\xi$. 
    }
    \label{Fig:ChangeGammaRad}
\end{figure*}
In Fig. \ref{Fig:ChangeGammaRad}, we have plotted the steady-state mean intra-cavity photon number and photon auto-correlation as functions of the pump rate for three different values of $\xi$, illustrating the effects of increasing or decreasing the spontaneous emission rate relative to the stimulated emission rate. We see several effects: For small pump rates, the reduction of the rate of spontaneous emission leads to a lower mean number of photons in the cavity as well as an increased value of the photon auto-correlation function to super-thermal values. Conversely, increasing the spontaneous emission rate leads to a larger mean intra-cavity photon number and sub-thermal photon statistics in the low-pump limit. In both cases, the results of the stochastic simulations follow the analytical results very well. In fact, we find that there is a simple relationship between the value of $\xi$ and the value of the photon auto-correlation function at low pump rates, namely
    \eql{g2_lowPump}{g^{(2)}(0) \rightarrow 1 + 1/\xi, \hskip1cm \text{for } \gamma_{p}\rightarrow 0.}
These effects occur in part because the alteration of the ratio $\xi$ from 1 leads to changes in one of the derived parameters $\beta = \gamma_{r}^{\rm sp}/(\gamma_{r}^{\rm sp} + \gamma_{bg})$ or $n_{th} = \gamma_{c}/\gamma_{r}^{\rm st}$. Note that the well-known result $g^{(2)}(0) = 2$ at low pump rates is restored in the case where the spontaneous and stimulated emission rates are equal, $\xi= \gamma_{r}^{\rm sp}/\gamma_{r}^{\rm st} = 1$.

From Fig. \ref{Fig:ChangeGammaRad} we see that decreasing the rate of spontaneous emission or increasing the rate of stimulated emission lead to the same qualitative effect, namely a reduced mean photon number and super-thermal photon statistics. Correspondingly, increasing the rate of spontaneous emission or decreasing the rate of stimulated emission both lead to a larger mean photon number and sub-thermal photon statistics. 
We can understand these effects qualitatively as follows: Less spontaneous emission will lead to fewer photons in the cavity in the low-pump regime where spontaneous emission dominates. This means that there will be more excited emitters which can be prompted to emit through stimulated emission by the few photons in the cavity, which in turn leads to a higher probability of several photons being present at the same time, hence a larger auto-correlation. Equivalently, an increased rate of stimulated events (stimulated emission, absorption) means that any emitted photons are more likely to be reabsorbed, giving fewer photons on average, while each photon has a larger probability to set off a stimulated emission from any excited emitter, resulting in a larger $g^{(2)}(0)$. The increased rate of stimulated emission is also the reason for a larger mean photon number in the high-pump limit seen in Fig. 4(b). The converse situation with a larger rate of spontaneous emission / lower rate of stimulated emission and absorption can be understood in a similar way.

\begin{figure*}[t]
    \centering
    \includegraphics[width=0.75\textwidth]{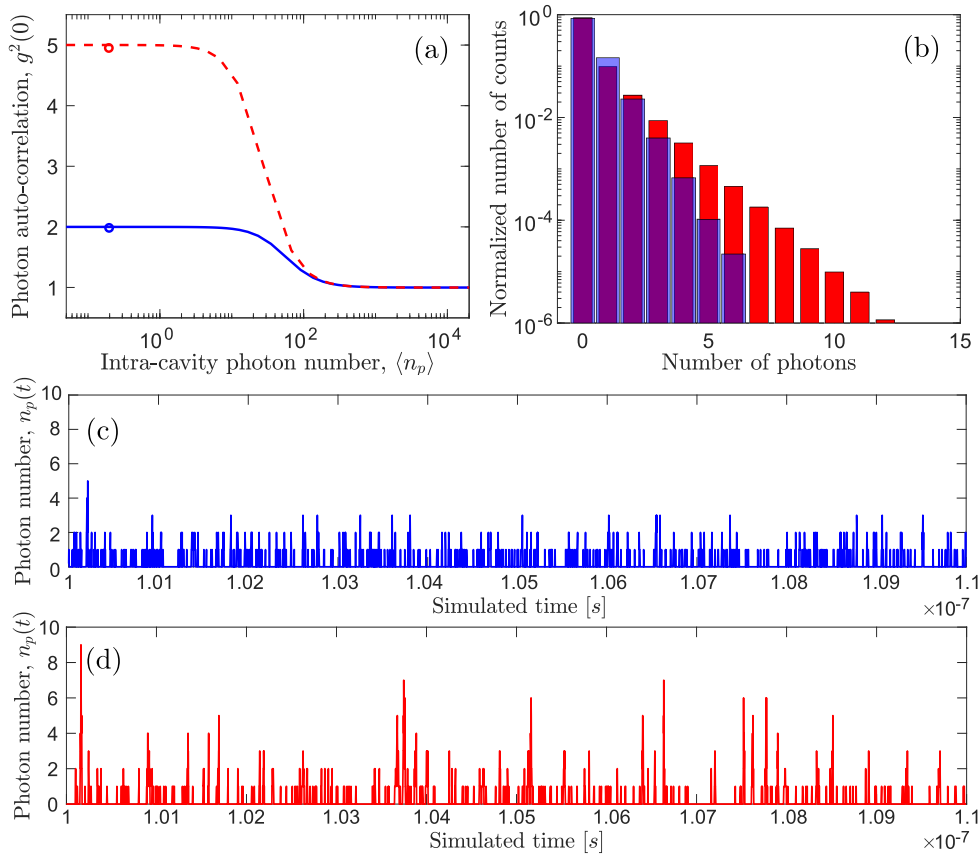}
    \caption{(a) Photon auto-correlation function versus mean photon number for $\xi=1$ [full blue line] and $\xi=1/4$ [dashed red line]. Circles indicate simulated results with steady-state mean photon number $\<n_p\> = 0.2$. (b) Simulated photon distribution for the two cases indicated by the circles in (a). (c,d) Time-resolved photon number for the two cases indicated in~(a) during 10~$ns$ of simulated time after reaching steady state. 
    }
    \label{Fig:ChangeGammaRad_TimeSeries}
\end{figure*}

From recent experiments and theories regarding nanolasers with collective correlations between the emitters, it is known that collective effects in small lasers can result in a reduced mean number of photons in the cavity and super-thermal photon statistics in the low-pump limit \cite{Leymann2015, Jahnke2016, Kreinberg2017}. Fig. \ref{Fig:ChangeGammaRad} shows that these phenomena also appear from the laser rate equations if one breaks the symmetry between the spontaneous and stimulated emission rates dictated by the Einstein relations. The effects are typically described theoretically by the emitter cross-correlations causing so-called excitation trapping at small pump rates \cite{Leymann2015,Jahnke2016}, meaning excitations in the system preferably reside in the emitters rather than the cavity mode. This effect, which is related to subradiance \cite{Dicke1954, GrossHaroche}, in turn gives rise to bursts of photon emission, similarly to what we described above in conjunction with the reduced rate of spontaneous emission. To see that such bursts of photons occur in our simulations for $\xi<1$, we have plotted in Fig. \ref{Fig:ChangeGammaRad_TimeSeries} the time-resolved photon number for two values of $\xi$. As shown in Fig. \ref{Fig:ChangeGammaRad_TimeSeries}(a,b), we examine the case where the steady-state mean photon number is the same, namely $\<n_p\> = 0.2$, while the photon distributions and statistics are either thermal or super-thermal. From (c,d) it is clear that the case where $\xi<1$, i.e. the rate of spontaneous emission is reduced, the photons often come in larger bursts, giving rise to the super-thermal statistics that we observe. This corresponds well with the theoretical descriptions given of the effects of emitter-emitter correlations in nanolasers, meaning an effectively reduced rate of spontaneous emission may provide a simple and intuitive way of further studying and understanding collective effects in nanolasers.

\section{Summary and discussion}\label{Sec:Summary}
In summary, we have presented Gillespie's First Reaction Method (FRM) in detail, and applied it to stochastic simulations of the traditional laser rate equations. We have shown that the FRM reduces computation times by around three orders of magnitude compared to earlier algorithms \cite{Mork2018}, while it also avoids potential numerical issues related to the size of time-increments and the ordering of events. Thus, the FRM is an efficient and exact algorithm for stochastic simulations of the laser rate equations.  We have used the FRM to examine the intra-cavity photon distribution as the laser transitions from chaotic to coherent emission, and we have shown that the numerical results for the photon statistics follow the analytical results exactly. We have also applied the FRM to altered laser rate equations, in which the symmetry between the spontaneous and stimulated emission rates was broken. We found that by reducing the rate of spontaneous emission, or increasing the rate of stimulated emission, both the numerical and analytical results show features that are also found in theories and experiments regarding lasers with strong emitter-emitter correlations. This suggests that there may be a way to use a slightly altered version of the well-known laser rate equations to describe collective effects in micro- and nanolasers. 

The FRM described here increases the efficiency of the stochastic simulations tremendously compared to methods like the FTI, while remaining exact and conceptually simple. One of the reasons why the \emph{exact} FRM algorithm is more efficient than the \emph{approximate} FTI algorithm for the laser rate equations is that there are relatively few event types and particles involved. Another reason is the conservative choice of time-increment size used in the FTI to obtain accurate results: In the FTI, and in the similar $\tau$-leap method \cite{Pahle2009,Gillespie2001}, faster computation times can always be obtained by choosing the time increments to be larger, but this naturally reduces the accuracy of the simulation. In the low-pump limit, where the photon statistics are very sensitive to the particle numbers, high accuracy is needed, implying small time-increments and hence long computation times. In the high-pump limit, where the particle numbers are large, it is possible that a less conservative choice of time-increment sizes could reduce the computation times of the FTI algorithm without significantly affecting the accuracy. Indeed, optimizing the size of the time-increment may lead to computation times for the FTI algorithm which are closer to those for the FRM algorithm in the high-pump limit, at a low cost to the accuracy. We leave an investigation of this to future work.

Another exact and conceptually simple algorithm for stochastic simulation is Gillespie's Direct Method \cite{Gillespie1976,Gillespie1977,Chusseau2003}, which is similar to the FRM, but requires only two random draws per iteration. This could potentially increase the efficiency of the simulation further, but due to the relatively low number of event types in the rate equations, the difference in computation time is expected to be minor. We performed a few simple simulation tests using Gillespie's Direct Method, and for the simulated times needed to obtain reliable statistics, the preliminary results showed practically no difference in computation time compared to the FRM. Both the FRM and the Direct Method have been expanded upon, leading to even more efficient algorithms \cite{Gibson2000,Pahle2009}. However, in most cases the increase in efficiency happens at the cost of the conceptual simplicity. For instance, the Next Reaction Method introduced in \cite{Gibson2000} reduces the computation time by storing and reusing the tentative times $\tau_{\mu}$ and only updating certain rates $a_{\mu}$, but it requires the introduction of additional concepts like dependency graphs and indexed priority queues. Since the rate equations for nanolasers deals with relatively few particles and event types, this added complexity would likely lead to fairly small improvements in efficiency. However, such algorithms could be implemented in stochastic simulations of the laser rate equations in future work.

Applying the FRM to stochastic simulations of the laser rate equations leads to short computation times, which makes it possible to experiment more easily with changes in system parameters and even alterations of the laser model. 
Using the FRM, one could potentially study other characteristics of the laser as well, like the emission spectrum and linewidth, and the response to different types of pumping. Additionally, if the method can be suitably modified, it can perhaps be used to further examine the effects of inter-emitter coupling in nanolasers through simulations using an extended set of rate equations \cite{ProtsenkoArxiv}. In conclusion, the FRM is efficient, robust and versatile, and we hope that this detailed description of the algorithm and how to apply it to the laser rate equations will lead to more new research and insights in the future.

\appendix

\section{Derivation of the FTI and FRM algorithms }\label{App:Algorithms}
In this appendix, we will give a slightly more detailed description of why the FTI and FRM algorithms have the specific forms described in the main text. We will describe in some detail the probability theoretical considerations that go into the derivation of the algorithms, though the full mathematical description is beyond the scope of this paper; see instead e.g. \cite{Gillespie1976}.

As mentioned in Sec. \ref{Sec:StochSim}, our aim is to simulate the interaction of a collection of discrete photons and emitters in a single-mode optical cavity, whose dynamics are governed by the set of rate equations \eref{RateEqs_p}-\eref{RateEqs_e}. To perform these stochastic simulations, we make a fundamental assumption common to the stochastic formulation of chemical kinetics \cite{Gillespie1976}: There exist a set of constants $c_{\mu}$ for each type of event $\mu$, which depend only on the physical properties of the system (not particle numbers), such that
    \ali{c_{\mu} dt =& \text{ Average probability, to first order in  }\nonumber \\
    & \text{ $dt$, that one particular combination of } \nonumber \\
    & \text{ photons and/or emitters  will undergo} \label{c_mu}\\
    &\text{ an event of type $\mu$ during the next } \nonumber \\
    &\text{ time increment } dt. \nonumber
    }
For instance, the probability that a particular photon in the cavity mode will be lost through the cavity mirrors during a time increment $dt$, averaged over all photons in the cavity mode, is $\gamma_{c} dt + o(dt)$, where $o(dt)$ are extra terms satisfying $o(dt)/dt \rightarrow 0$ as $dt\rightarrow 0$; In other words, $c_{c} = \gamma_{c}$. Likewise, the probability that a particular photon in the cavity mode will cause a specific excited emitter to emit in a stimulated emission event during $dt$ is $\gamma_{r} dt + o(dt)$, so $c_{st} = \gamma_{r}$. Additionally, if at time $t$ the numbers of photons and excited emitters are $n_{p}(t)$ and $n_{e}(t)$, respectively, we can define the functions $h_{\mu}$ as
    \ali{h_{\mu} =& \text{ Number of distinct combinations of photons } \nonumber \\
    & \text{ and/or emitters that can undergo an event} \label{h_mu}\\
    &\text{ of type $\mu$, given that the current particle} \nonumber \\
    &\text{ numbers are }  (n_{p}(t) , n_{e}(t)). \nonumber
    }
For instance, the number of photons which could potentially leak out of the cavity is the current photon number, i.e. $h_{c} = n_{p}(t)$; and since each of the $n_{p}(t)$ photons can prompt a stimulated emission event from each of the $n_{e}(t)$ excited emitters, we take $h_{st} = n_{e}(t) n_{p}(t)$. Combining the two quantities defined in eqs.  \eref{c_mu} and \eref{h_mu}, we can define the rates, or propensity functions, $a_{\mu}$ as
    \ali{a_{\mu} dt =& \ h_{\mu} c_{\mu} dt \nonumber \\
    =& \text{ Probability, to first order in $dt$, }\nonumber \\
    & \text{  that an event of type $\mu$ will } \label{a_mu} \\
    & \text{ occur during the next time } \nonumber\\
    &\text{ increment } dt. \nonumber
    }
In \cite{Mork2018}, the stochastic simulations are carried out by choosing a fixed time increment $dt$ and using the quantities defined above to determine how many of each type of event $\mu$ that happen at every time step: It is suggested that the total number of events of type $\mu$ happening during the time increment $dt$ follows a binomial distribution, $\text{Binom}(h_{\mu} , c_{\mu}dt)$, based on the idea that each individual event which happens can be seen as a success in a Bernoulli trial, i.e. a yes/no-experiment, whose probability of success is $c_{\mu} dt$ (Of course this assumes that $dt$ is small enough that $c_{\mu}dt\leq 1$). As an example, the number of photons lost through the cavity mirrors between $t$ and $t+dt$ is taken to be a random integer drawn from the binomial distribution $\text{Binom}(n_{p}(t) , \gamma_{c} dt)$, since the average probability of losing one photon is $\gamma_{c} dt$ and there are $n_{p}(t)$ "tries". In order to speed up the simulations, a further assumption is made in \cite{Mork2018}, namely that the time increment $dt$ may be chosen sufficiently small that the binomial distributions $\text{Binom}(h_{\mu} , c_{\mu}dt)$ may be replaced by Poisson distributions $\text{Poiss}(h_{\mu}c_{\mu} dt) = \text{Poiss}(a_{\mu} dt)$. This removes the upper bound on the integers that may be drawn, introducing the risk that e.g. the number drawn for how many photons are lost through the cavity mirrors is greater than the number of photons currently in the cavity mode, but by choosing a sufficiently small time increment $dt$, this risk is minimal.

While the replacement of binomial distributions by Poisson distributions significantly reduces computation times for the algorithm of \cite{Mork2018}, the simulation must still run for several hours to obtain satisfactory photon statistics for all pump values. Implementing a more efficient algorithm is therefore highly desirable, and to do this we can change the fundamental viewpoint in the simulation: Instead of determining how many events of each type occur during a fixed time increment, we can try to determine the length of time until \emph{some} event occurs and then determine which type of event occurs. There are several ways of doing this, but Gillespie's First Reaction Method is based on the following idea: For each $\mu$ we can find a tentative time $\tau_{\mu}$ until an event of type $\mu$ would occur, assuming no other event occurs before, and then we choose the smallest of these tentative times, $\tau_{\mu_0}$, as the time until an event actually occurs. The corresponding event type, $\mu_{0}$, is chosen as the event type that happens. To generate the random $\tau_{\mu}$'s, we should consider the probabilities at time $t$ for an event of type $\mu$ to happen between times $t+\tau$ and $t+\tau +dt$. We may compute such a probability as the product of the probability that no event of type $\mu$ happens between $t$ and $t+\tau$, given by
    \eql{ProbEventMu}{P_{0}(\tau) = \exp(- a_{\mu} \tau), }
and the probability $a_{\mu} dt$ that an event of type $\mu$ does occur during an interval of length $dt$ \cite{Gillespie1976}. In other words, the probabilities that we are looking for are of the form
    \eql{ExpDistMu}{P_{\mu}(\tau) \;dt = a_{\mu}  \exp(- a_{\mu} \tau)\;dt.}
To generate the random tentative times $\tau_{\mu}$, we can therefore draw random numbers from the distributions $P_{\mu}(\tau)$, which are essentially exponential distributions with parameters $a_{\mu}$. 

\section{Funding}
Villum Fonden (8692); Danish National Research Foundation, grant number DNRF147.

\section{Disclosures}
The authors declare no conflicts of interest.

\section{Acknowledgments}
The authors wish to thank I.E. Protsenko and A.V. Uskov for stimulating discussions regarding collective effects in nanolasers. J.M. wishes to thank G.L. Lippi for helpful discussions regarding stochastic simulations.

\bibliography{Stochastic_Bib}

\end{document}